\documentclass[11pt]{article}

\usepackage{fullpage}
\usepackage{multicol}
\usepackage{times}
\usepackage{amsmath}
\usepackage{amsthm}
\usepackage{amssymb}
\usepackage{enumerate}

\usepackage{algorithm}
\usepackage{algorithmic}

\newtheorem{Theorem}{Theorem}

\newtheoremstyle{definition}{}{}
     {\rmfamily}
     {}
     {\bfseries}
     {.}
     { }
     {}
\theoremstyle{definition}
\newtheorem{Definition}{Definition}

\newenvironment{Proof}[0]{\begin{proof}}{\end{proof}} 

\newcommand{\Ar}{\mathcal{A}}

\newcommand{\Cr}{\mathcal{C}}

\newcommand{\Nm}{\mathbb{N}}

\renewcommand{\epsilon}{\varepsilon}
\newcommand{\toset}{\rightarrow}

\newcommand{\oracle}{\mathcal{O}}
\newcommand{\argmin}{\mathrm{argmin}}

\newcommand{\steepone}{\textsf{Randomized Steepest Descent}}
\newcommand{\steeptwo}{\textsf{Quantum Steepest Descent}}

\begin{document}
\date{}
\title{Enhanced algorithms for Local Search}
\author{Yves F. Verhoeven\thanks{Laboratoire de Recherche en Informatique, B\^atiment 490,
Universit\'e Paris-Sud, 91405 Orsay Cedex, France and \'Ecole Nationale Sup\'erieure
des T\'el\'ecommunications, 46 rue Barrault, 75013 Paris, France.
E-mail:~{\tt yves.verho\hspace{-20cm}.\hspace{19.8cm}even@normalesup.org}.
The research was supported by the EU 5th framework program RESQ IST-2001-37559,
and by the ACI CR 2002-40 and ACI SI 2003-24 grants of the
French Research Ministry.}}
\maketitle

\begin{abstract}
Let $G=(V,E)$ be a finite graph, and $f:V\toset \Nm$ be any function. The Local
Search problem consists in finding a {\em local minimum of the function $f$ on
$G$}, that is a vertex $v$ such that $f(v)$ is not larger than the value of $f$
on the neighbors of $v$ in $G$. In this note, we first prove a separation
theorem slightly stronger than the one of Gilbert, Hutchinson and Tarjan for
graphs of constant genus. This result allows us to enhance a previously known
deterministic algorithm for Local Search with query complexity
$O(\log n)\cdot d+O(\sqrt{g})\cdot\sqrt{n}$, so that we obtain a deterministic
query complexity of $d+O(\sqrt{g})\cdot\sqrt{n}$, where $n$ is the size of $G$,
$d$ is its maximum degree, and $g$ is its genus. We also give a quantum version
of our algorithm, whose query complexity is of
$O(\sqrt{d})+O(\sqrt[4]{g})\cdot\sqrt[4]{n}\log\log n$. Our deterministic and
quantum algorithms have query complexities respectively smaller than the
algorithm \steepone{} of Aldous and \steeptwo{} of Aaronson for large classes of
graphs, including graphs of bounded genus and planar graphs. Independently from
this work, Zhang has recently given a quantum algorithm which finds a local minimum on
the planar grid over $\{1,\ldots,\sqrt{n}\}^2$ using
$O(\sqrt[4]{n}(\log\log n)^2)$ queries. Our quantum algorithm can be viewed as a
strongly generalized, and slightly enhanced version of this algorithm.
\end{abstract}

\begin{multicols}{2}
\section{Introduction}
The Local Search problem consists in finding a {\em local minimum of the function $f$ on $G$},
that is a vertex $v$ such that $f(v)$ is not larger than the value of $f$ on the neighbors
of $v$ in $G$. Obviously, such a vertex always exists, as a global minimum satisfies this constraint.
Another easy argument shows how to find such a vertex : make a walk
over vertices such that at each step the next vertex is the neighbor of the current vertex which has
the smallest value; the walk will stop in a local minimum.
Such a walk is called a {\em steepest descent}.
Steepest descents are the basis of several approaches to efficiently find a local minimum.

The Local Search problem has been previously studied and there is already a large literature
on its complexity. Its structural complexity, where the function and the graph are
given as an input to a Turing machine, was studied in~\cite{pls_class, pap_schaf_yann},
and its query complexity, where
the graph is known but the values of $f$ are accessed through an oracle, was investigated
in~\cite{Llewellyn_Tovey_1,Llewellyn_Tovey_2,Aldous_min_search,Aaronson_min_search,Miklos_Mario,zhang}.
We focus on the query complexity, which complexity is obviously at most
the size of the graph. Our query model is the standard one;
see Section~\ref{sec:query} for precise definitions.

The deterministic query complexity of Local Search on a graph $G$ of size $n$, maximum degree $d$ and
genus $g$~(for a definition of the genus, see for instance~\cite{modern_graph_theory}),
is intimately connected to the size of {\em separators} of $G$:

\begin{Definition}
A {\em separator for $G$} is a subset of $V$ whose removal leaves no connected component with more
than $2n/3$ vertices.
\end{Definition}

In~\cite{Llewellyn_Tovey_1}, a deterministic query algorithm was exhibited, which works
using a sub-linear number of queries for large classes of graphs: a local minimum can be found on the
graph $G$ using $O(\log n)\cdot d+O(g)\cdot\sqrt{n}$ queries.
It is based on the recursive use of separators for smaller and smaller subgraphs of $G$,
and their complexity analysis relies on the following result:

\begin{Theorem}[Gilbert, Hutchinson, Tarjan~\cite{separator_genus}]
\label{th:separation_genus}
The graph $G$ has a separator of size at most $6\sqrt{gn}+2\sqrt{2n}+1$.
\end{Theorem}

In the randomized and quantum query models, the situation is quite different, as
the size of separators of $G$. Also, the only known sub-linear query algorithm for general graphs is a randomized algorithm,
that we call \steepone{},
was exhibited by Aldous~\cite{Aldous_min_search}, and has a query complexity $\Theta(\sqrt{nd})$.
The idea of this algorithm is to choose $\sqrt{nd}$ vertices at random, query their values, start a steepest
descent from the vertex with smallest value for at most $\sqrt{nd}$ steps, and to return the last visited vertex.
This idea was later refined by Aaronson~\cite{Aaronson_min_search}
to give a sub-linear quantum query algorithm for general graphs, that we call \steeptwo{}, using $\Theta(n^{1/3} d^{1/6})$ queries.

On the side of lower-bounds, it follows from~\cite{Llewellyn_Tovey_1} that the size of a smallest separator is a lower-bound on
the deterministic complexity of Local Search. Also, from~\cite{Aaronson_min_search},
we know that $d$ is a lower-bound on the deterministic query complexity, $\Omega(d)$ a lower-bound
for the randomized query complexity, and $\Omega(\sqrt{d})$ a lower-bound
for the quantum query complexity.

\section{Results}
In this note, we first improve Theorem~\ref{th:separation_genus} in Section~\ref{sec:separation},
to obtain the following slightly stronger separation theorem:

\begin{Theorem}[strong separation for graphs of genus $g$]
\label{th:strong_separation_genus}
Assume $n\geq 3$. There exists a separator $C$ for $G$ 
such that $C$ contains no more than $(6+2\sqrt{2}-12/n+6\sqrt{g}+4g/n+1/\sqrt{n})\cdot\sqrt{n}$ vertices,
and the subgraph induced on $G$ by $V\setminus C$ has maximal degree at most $\sqrt{n}$.
\end{Theorem}

As a result it allows us to enhance, in Section~\ref{paragraph:upper_bound},
the deterministic algorithm of Llewellyn, Tovey and Trick~\cite{Llewellyn_Tovey_1}
whose complexity is of $O(\log n)\cdot d+O(\sqrt{g})\cdot\sqrt{n}$. 
We also derive a quantum algorithm from it.
More precisely, we obtain the following result:

\begin{Theorem}
\label{th:algo_presentation}
There exists a deterministic and a quantum query algorithms that find a local minimum of $f$ on $G$
using respectively $d+O(\sqrt{g})\cdot\sqrt{n}$ and $O(\sqrt{d})+O(\sqrt[4]{g})\cdot\sqrt[4]{n}\log\log n$
queries.
\end{Theorem}

Our deterministic and quantum algorithms have smaller query complexities than the
respective algorithms \steepone{} of Aldous~\cite{Aldous_min_search} and of \steeptwo{}
of Aaronson~\cite{Aaronson_min_search} for large classes of graphs,
including graphs of bounded genus and planar graphs.
We analyze this in detail in Section~\ref{sec:comparison}.

Independently from this work, Zhang~\cite{zhang} has recently given an algorithm
which finds a local minimum on the planar grid over $\{1,\ldots,\sqrt{n}\}^2$
using $O(\sqrt[4]{n}(\log\log n)^2)$ queries. Our quantum algorithm can be viewed as
a strongly generalized, and slightly enhanced version of this algorithm.

\section{Preliminaries}
\subsection{Notations}
We denote by $\log n$ the natural logarithm of $n$, and
for every positive real number $b$ we denote by $\log_b n$ the logarithm of $n$ in base $b$.
If $G$ is a graph and $v$ is any vertex of $G$, we denote by $\partial_G(v)$ the set
of neighbors of $v$ in $G$.

\subsection{Query complexity}
\label{sec:query}
In the query model of computation we count only queries made by the algorithm, but
all other computations are free. The state of the computation is represented by three
registers, the query register $i \in \{1,\ldots,n\}$, the answer register $a \in \Sigma$, and the 
work register $z\in W$, where $\Sigma$ and $W$ are finite sets.
The computation takes place in the vector space spanned by all
basis states $|i\rangle|a\rangle|z\rangle$.
In the {\em quantum query model} introduced by Beals, Buhrman,
Cleve, Mosca and de Wolf~\cite{BBC} the state of the computation is a complex
combination of all basis states which has unit length for the norm $\ell_2$,
and the allowed operations on the state of the computation
are all isometric operators for the $\ell_2$ norm acting over the computation space.
In the randomized model, the state of the computation
is a non-negative real combination of all basis states of unit length
for the norm $\ell_1$, and the allowed operations on the state of the computation
are all isometric operators for the norm $\ell_1$ acting over the computation space.
In the deterministic model, the state of the computation
is always one of the basis states, and the allowed operations
are all operators mapping a basis state to another basis state.

Assume that $x\in\Sigma^n$ is the input of the problem which can be accessed only through
the oracle.
The query operation $\oracle_x$ is the permutation which maps the basis state
$|i\rangle|a\rangle|z\rangle$  into the state $|i\rangle|(a+x_i) \bmod |\Sigma|\rangle|z\rangle$
(here we identify $\Sigma$ with the residue classes $\bmod |\Sigma|$).
Non-query operations are independent of $x$.
A {\em $k$-query algorithm} is a sequence of $(k+1)$ operations
$(U_0, U_1, \ldots , U_k)$ where $U_i$ is an allowed operation in the chosen model of computation.
Initially the state of the computation is set to some
fixed value $|0\rangle|0\rangle|0\rangle$, and then the sequence of operations
$U_0, \oracle_x, U_1, \oracle_x, \ldots, U_{k-1}, \oracle_x, U_k$ is applied.
The final state is denoted by $\Phi$.

The output in the quantum model is an element $z\in W$ that
appears with probability equal to the square
of the $\ell_2$ norm of the orthogonal projection of $\Phi$ over
the vector space $V$ spanned by $\{|i\rangle|a\rangle|z\rangle \,|\,
i \in \{1,\ldots,n\},\, a \in \Sigma\}$.
The output in the randomized model is an element $z\in W$ that
appears with probability equal to the $\ell_1$ norm of the orthogonal projection of $\Phi$ over
the vector space $V$.
The output in the deterministic model is the element $z\in W$
such that there exist $i$ and $a$ with $\Phi=|i\rangle|a\rangle|z\rangle$.

Assume that $R\subseteq \Sigma^n\times W$ is a total relation~({\em i.e.} for every
$x\in\Sigma^n$ there exists $z\in W$ such that $(x,z)\in R$) that we 
want to compute. A quantum or randomized algorithm computes~(with two-sided error)
$R$ if its output yield some $z \in W$ such that $(x,z) \in R$ with probability 
at least $2/3$. A deterministic algorithm computes $R$ if its output yield
some $z \in W$ such that $(x,z) \in R$.

Then the query complexity of a relation $R$ in a model of
computation (deterministic, randomized or quantum) is the smallest
$k$ for which there exists a $k$-query algorithm, in that model
of computation, which computes $R$.

\section{Tools}
In this section, we recall and prove the results that we need in order to design the
algorithms of Section~\ref{paragraph:upper_bound}.

\subsection{Separation in graphs of higher genus}
\label{sec:separation}
We first recall the following well-known theorem for graphs of higher
genus~(see for instance~\cite{modern_graph_theory})~:
\begin{Theorem}
\label{th:genus_inequality}
Any $n$-vertex graph of genus $g$ with $n\geq 3$ contains no more than $3n-6+2g$ edges.
\end{Theorem}

This result, together with Theorem~\ref{th:separation_genus}, allows us to prove Theorem~\ref{th:strong_separation_genus}.

\begin{Proof}[Proof of Theorem~\ref{th:strong_separation_genus}]
Theorem~\ref{th:separation_genus} shows that it is possible to find a separator $C'$ for $G$ such
that $C'$ has size at most $6\sqrt{gn}+2\sqrt{2n}+1$.
Let $B$ be the set of all vertices of degree greater than $\sqrt{n}$. Using
Theorem~\ref{th:genus_inequality}, we have
\begin{equation*}
|B|\cdot\sqrt{n} \leq \sum_{v\in V} \mathrm{d}(v)=2\cdot|E|\leq 6n-12+4g.
\end{equation*}
The set $C=C'\cup B$ being a superset of a separator of $G$ is also
a separator of $G$. From the definition of $B$, the subgraph induced on $G$ by $V\setminus C$
obviously has maximal degree at most $\sqrt{n}$.
\end{Proof}

The genus of a graph being in $O(n^2)$, the asymptotic inequality $g/n=O(\sqrt{g})$ holds and therefore
Theorem~\ref{th:strong_separation_genus} can be interpreted as stating the existence of
a particular $O(\sqrt{g})\cdot\sqrt{n}$ separator $G$.

\subsection{Minimum-finding algorithms}
\label{sec:min-finding}

In this paragraph, we recall results about the query complexity of finding
the minimum value of a function on a set.

Let $n$ be a positive integer, $S$ be a set of cardinality $n$,
$g:S\toset\Nm$ be a function and $\oracle_g$ be an oracle for $g$.

\begin{Definition}
Let $\epsilon<1$ be any positive real number.
If $\Ar$ is a randomized algorithm that outputs the minimum value of the function $g$
on $S$ with probability at least $1-\epsilon>0$, then we denote by
$\argmin_\Ar\{g(s) \,|\, s\in S\}$ the random variable equal to its output.
\end{Definition}

It is obvious that, for every deterministic algorithm $\Ar$, computing
$\argmin_\Ar\{g(s) \,|\, s\in S\}$ requires querying all $n$ values of $g$ to $\oracle_g$.
It is natural that, for every randomized algorithm $\Ar$, computing
$\argmin_\Ar\{g(s) \,|\, s\in S\}$ requires querying $\Omega(n)$ values of $g$ to $\oracle_g$.
It is more surprising that much less queries are needed when quantum queries are allowed:

\begin{Theorem}[D{\"u}rr, H{\o}yer~\cite{Durr_Hoyer}]
There exists a quantum algorithm which finds the minimum value of
$g$ with probability at least $1/2$, using $O(\sqrt{n})$ quantum queries to $\oracle_g$.
\end{Theorem}

Amplification of the probability of success of the algorithm of D{\"u}rr and H{\o}yer
can be obtained by running the algorithm several
times, and then taking the minimum value of all the values that have been returned by each
repetition of the algorithm. After $k$ repetitions, the probability of having found the minimum
is at least $1-2^{-k}$. In particular, for every positive real number $\epsilon<1$,
there exists a quantum algorithm $\Ar$ computing $\argmin_\Ar\{g(s) \,|\, s\in S\}$ with probability at least $1-\epsilon$
using $O(\sqrt{n}\log(1/\epsilon))$ quantum queries.

\section{Enhanced algorithms for Local Search}
\label{paragraph:upper_bound}
In this paragraph, we prove Theorem~\ref{th:algo_presentation}.
The proof will be in two steps: in Theorem~\ref{th:correction} we will prove the correction of our algorithms,
and in Theorem~\ref{th:analysis} we will prove their complexity.

The basic procedure of our algorithms follows the lines of the algorithms of Llewellyn,
Tovey and Trick of~\cite{Llewellyn_Tovey_1} and Santha and Szegedy~\cite{Miklos_Mario}.
It is given in Algorithm~\ref{algo:minimum}.
The main idea is to adopt a divide-and-conquer approach: the graph is split into
connected components of small size by removing a separator; then, querying
the values of the vertices in and close to that separator make it possible to
find one of these connected components in which there is a local minimum of $f$ on $G$.

Notice that neither the way the separators are chosen, or how the minimum-finding
algorithms $\Ar_i$ work for integers $i\geq 1$, are specified in Algorithm~\ref{algo:minimum}.
Our algorithms consist in using the procedure described in Algorithm~\ref{algo:minimum} with
the following specific choices:
\begin{itemize}
\item a separator $C$ of a graph $G'$ will be chosen according to Theorem~\ref{th:strong_separation_genus}
  if $G'$ has more than two vertices, and $C$ contains all vertices of $G'$ otherwise,
\item the minimum-finding algorithm $\Ar_i$ will behave as follows when requested to minimize the function
  $f$ over a set $S\subseteq V$ of vertices: in the deterministic case,
  the local minimum of $f$ is always found by exhaustive search\footnote{One should observe that
  it is important at this point that $f$ takes distinct values on distinct vertices. This can be
  assumed, as one could for instance put a total order $\prec$ on $V$, and minimize the function
  $g:v\mapsto (f(v),v)$ according to the lexicographic order induced on $\Nm\times V$ by $<$ and $\prec$,
  instead of minimizing $f$. The function $g$ takes distinct values on distinct vertices.}.
  In the quantum cases, if the set $S$ has size at most $3$, then
  the minimum value of $f$ on $S$ is also found using exhaustive search. Otherwise,
  the output is the one found by the last measurement at the end of the quantum procedure
  described in paragraph~\ref{sec:min-finding}; moreover, we request that the minimum-finding
  algorithm $\Ar_1$ has error probability $1/12$ using $O(\sqrt{|S|})$ queries, and $\Ar_i$ has error probability
  $1/12\log_{3/2} n$ using $O(\sqrt{|S|}\log\log n)$ queries, for $i\neq 1$.
\end{itemize}

\end{multicols}
\begin{algorithm}[h]
\caption{Procedure for finding a local minimum of a function $F:V\toset\Nm$ on a graph $G=(V,E)$, using separators.}
\begin{algorithmic}
\STATE $i:=0$, $G^{(0)}:=G$, $v^{(0)}:=\text{any vertex of } G$, $\mathsf{output}:=\emptyset$.
\WHILE{$\mathsf{output}=\emptyset$}
    \STATE $i:=i+1$.
    \STATE Create a separator $C^{(i)}$ for $G^{(i-1)}$.
    \STATE $m^{(i)}:=\argmin_{\Ar_i}\{ f(v) \,|\, v\in C^{(i)}\}$.
    \STATE $z^{(i)}:=\argmin_{\Ar_i}\{ f(v) \,|\, v\in\partial_{G^{(i-1)}}(m^{(i)})\}$.
    \STATE $v^{(i)}:=\argmin_{\Ar_i}\{ f(v) \,|\, v\in\{v^{(i-1)},m^{(i)},z^{(i)}\}\}$.
    \IF{$v^{(i)}=m^{(i)}$}
      \STATE $\mathsf{output}:=\{v^{(i)}\}$.
    \ELSE
      \STATE $V^{(i)}:=\text{the connected component of } V^{(i-1)}\setminus C^{(i)}\text{ that contains } v^{(i)}$.
      \STATE $G^{(i)}:=G[V^{(i)}]$.
    \ENDIF
\ENDWHILE
\STATE Return $\mathsf{output}$.
\end{algorithmic}
\label{algo:minimum}
\end{algorithm}
\begin{multicols}{2}

\begin{Theorem}
\label{th:correction}
With our choice of minimum-finding algorithm, Algorithm~\ref{algo:minimum} always returns a local minimum
in the deterministic case, and returns a local minimum with probability at least $2/3$ in the quantum case.
\end{Theorem}

\begin{Proof}
Let $j$ be the largest value of the variable $i$ for a run of the algorithm.
First, an easy inductions shows that
for every iteration $i\leq j$ of the main loop, and every $v\in V^{(i)}$ we have
\begin{equation*}
\partial_{G}(v)\subseteq \partial_{G^{(i)}}(v)\cup C^{(1)}\cup C^{(2)}\cup\cdots\cup C^{(i)}.
\end{equation*}
So, to prove that $f$ is minimized on $v^{(j)}$, one must only prove that $f(v^{(j)})$ is not larger than
$\min\{f(v) \,|\, v\in \partial_{G^{(j)}}(v)\cup C^{(1)}\cup C^{(2)}\cup\cdots\cup C^{(j)}\}$.

If, during the run of the algorithm, the calls to the algorithms $\Ar_i$, for $1\leq i\leq j$,
have always successfully returned elements minimizing $f$, then
for every positive integer $i\leq j$ we have $f(v^{(i)})\leq\min\{f(v^{(i-1)}),f(m^{(i)}),f(z^{(i)})\}$. Therefore, an easy induction
shows that $f(v^{(i)})\leq\min\{f(v) \,|\, v\in C^{(k)} \}$, for every positive integers $k\leq i\leq j$. Moreover,
the equality $v^{(j)}=m^{(j)}$ implies $f(v^{(j)})\leq\min\{f(v) \,|\, v\in\partial_{G^{(j)}}(v^{(j)}) \}$. So, if
$\Ar_i$ never failed to find a minimizing element,
then the criterion given in the previous paragraph shows that $v^{(j)}$ is a local minimum.

In the deterministic case, the algorithms $\Ar_i$, for $1\leq i\leq j$,
always return an element minimizing $f$, and therefore
Algorithm~\ref{algo:minimum} always returns a local minimum.

In the quantum case, a call to $\Ar_i$
returns an element minimizing $f$ with error probability at most $1/12$ for $i=1$, and at most $1-1/4\log_{3/2} n$
for $1<i\leq j$. The set $C^{(i)}$ being a separator of $G^{(i-1)}$ for every positive integer $i\leq j$,
we have $|V^{(i)}|\leq 2|V^{(i-1)}|/3$. This implies
that $j\leq \log_{3/2} n$, and the probability that $\Ar_i$ did not return an element minimizing $f$
at some point is at most $2\cdot 1/12+2\cdot \log_{3/2} n/(12\log_{3/2} n)=1/3$. So, in the quantum case,
Algorithm~\ref{algo:minimum} returns a local minimum with probability at least $2/3$.
\end{Proof}

\begin{Theorem}
\label{th:analysis}
With our choices of separators, Algorithm~\ref{algo:minimum} has a deterministic query complexity
at most $d+O(\sqrt{g})\cdot\sqrt{n}$, and a quantum query complexity at most
$O(\sqrt{d})+O(\sqrt[4]{g})\cdot \sqrt[4]{n}\log\log n$.
\end{Theorem}

\begin{Proof}
Again, let $j$ be the largest value of the variable $i$ for a run of the algorithm.
Let us denote by $\Cr_{\Ar_i}(s)$ the number of queries made by the minimum-finding
algorithm $\Ar_i$ on a set of size $s$,
and by $L^i(n,d)$ the number of queries that are made in the
$i$-th iteration of the main loop of our algorithm
on a graph $G'$ that has $n$ vertices and is of maximum degree $d$.
We denote also by $d_i$ the maximum degree of $|G^{(i)}|$, for a non-negative integer $i\leq j$.
Analysis of the main loop of Algorithm~\ref{algo:minimum} gives, for
every positive integer $i\leq j$,
\begin{equation*}
L^i(n,d)\leq \Cr_{\Ar_i}(|C^{(i)}|)+\Cr_{\Ar_i}(d_{i-1})+3.
\end{equation*}
Let us denote by $T_\gamma^i(\alpha,\beta)$ the number of queries made
by our algorithm in the main loop between its $i$-th iteration and
the end of the algorithm if $i<j$ and $0$ if $i\geq j$, on an input graph which has $\alpha$ vertices, is of maximum degree $\beta$
and has genus at most $\gamma$.
Theorem~\ref{th:strong_separation_genus} ensures that for every positive integer $i\leq j$ we have $|V^{(i)}|\leq 2|V^{(i-1)}|/3$,
and $d_i\leq \sqrt{|V^{(i-1)}|}$. Moreover, the genus of be $G^{(i)}$ is not larger
than the genus of $G^{(i-1)}$. So, by induction we have $|V^{(i)}|\leq (2/3)^i n$, and the genus of $|G^{(i)}|$ is at most $g$.
Therefore, for every integer $1\leq i\leq j$ we have $|C^{(i)}|\leq O(\sqrt{g})\cdot\sqrt{(2/3)^{i-1} n}$, $d_0=d$ and
$d_i\leq\sqrt{(2/3)^{i-1}n}$.
This leads to the following equations:
\begin{align*}
& T_g^1(n,d)\\
& \qquad\leq L^1(n,d)+T_g^2(n,d) \\
& \qquad\leq \Cr_{\Ar_1}(O(\sqrt{g})\cdot\sqrt{n})+\Cr_{\Ar_1}(d)+3+T_g^2(n,d),
\end{align*}
and for every $i\in\{2,\ldots,\lfloor\log_{3/2} n\rfloor-1\}$,
\begin{align*}
  & T_g^i(n,d) \\ 
  & \qquad\leq L^i(n,d)+T_g^{i+1}(n,d) \\
  & \qquad\leq \Cr_{\Ar_i}\left(O(\sqrt{g})\cdot\sqrt{(2/3)^{i-1} n}\right)+ \\
  & \qquad\qquad\Cr_{\Ar_i}\left(\sqrt{(2/3)^{i-2}n}\right)+3+ T_g^{i+1}(n,d).
\end{align*}
In the deterministic case we have $\Cr_{\Ar_i}(k)=k$ for all positive integers $k$ and $i$, and in the
quantum case we have, for all positive integer $k$, $\Cr_{\Ar_i}(k)=O(\sqrt{k})$ when $i=1$, and
$\Cr_{\Ar_i}(k)=O(\sqrt{k}\log\log n)$ when $i\neq 1$.
So, in the deterministic case, summing all the previous inequalities gives
\begin{multline*}
T_g^1(n,d) \leq O(\sqrt{g})\cdot\sqrt{n}\cdot\sum_{i=0}^\infty \sqrt{2/3}^i+d+ \\
  \sqrt{n}\cdot\sum_{i=0}^\infty \sqrt{2/3}^i+3\log_{3/2} n+ \\
  T_g^{\lfloor\log_{3/2} n\rfloor}(n,d),
\end{multline*}
which shows $T_g^1(n,d)=d+O(\sqrt{g})\cdot\sqrt{n}$, as $T_g^{\lfloor\log_{3/2} n\rfloor}(n,d)=O(1)$,
and the query complexity of our deterministic algorithm is $T_g^1(n,d)=d+O(\sqrt{g})\cdot\sqrt{n}$.
In the quantum case, it gives
\begin{multline*}
T_g^1(n,d) \leq 
  O(\sqrt[4]{g})\cdot\sqrt[4]{n}\log\log n\cdot\sum_{i=0}^\infty \sqrt[4]{2/3}^i+\\
  O(\sqrt{d})+O(\sqrt[4]{n}\log\log n)\cdot\sum_{i=0}^\infty \sqrt[4]{2/3}^i+\\
  3\log_{3/2} n+T_g^{\lfloor\log_{3/2} n\rfloor}(n,d),
\end{multline*}
leading to a quantum query complexity $T_g^1(n,d)=O(\sqrt{d})+O(\sqrt[4]{g})\cdot\sqrt[4]{n}\log\log n$.
\end{Proof}

\section{Comparison with generic algorithms}
\label{sec:comparison}
Let us first compare the query complexity of our deterministic algorithm with the query complexity
of the algorithm \steepone{} of Aldous~\cite{Aldous_min_search}. The complexity of our
algorithm is $d+O(\sqrt{g})\cdot\sqrt{n}$, and the complexity of \steepone{} is $\Theta(\sqrt{nd})$.
As $d\leq n$, our algorithm performs as well as \steepone{}~(up to a constant speedup factor) as soon as $g=O(d)$,
and performs asymptotically better when $g=o(d)$. In particular, our deterministic algorithm has lower
query complexity than \steepone{} on classes of graphs with bounded genus,
which includes the class of planar graphs.

Let us now compare the query complexity of our quantum algorithm with the query complexity
of the algorithm \steeptwo{} of Aaronson~\cite{Aaronson_min_search}. The complexity of our
algorithm is $O(\sqrt{d})+O(\sqrt[4]{g})\cdot\sqrt[4]{n}\log\log n$, and the complexity of \steeptwo{}
is $\Theta(n^{1/3} d^{1/6})$. As $d\leq n$, we have $\sqrt{d}\leq n^{1/3} d^{1/6}$, and
our algorithm performs as well as \steeptwo{}~(up to a constant speedup factor)
as soon as $g^{1/2}\cdot n^{1/4}\log\log n=O(n^{1/3} d^{1/6})$, that is to say $g=O(n^{1/6}d^{1/3}/(\log\log n)^2)$.
This holds if $g=O(\sqrt{d}/(\log\log n)^2)$. Also, our quantum algorithm
performs asymptotically better when $g=o(n^{1/6}d^{1/3}/(\log\log n)^2)$, and therefore when
$g=o(\sqrt{d}/(\log\log n)^2)$. In particular, our quantum algorithm has lower
query complexity than \steeptwo{} on classes of
graphs with bounded genus, which includes the class of planar graphs.

In conclusion, the algorithms we have designed perform better than the known
generic algorithms for some classes of graphs, in particular planar graphs and graphs of constant genus,
both for classical (deterministic and randomized) computation, and for quantum computation.


\bibliographystyle{plain}
\bibliography{planar_pls}

\begin{thebibliography}{10}

\bibitem{Aaronson_min_search}
Scott Aaronson.
\newblock Lower bounds for local search by quantum arguments.
\newblock In {\em Proceedings of the thirty-sixth annual ACM symposium on
  Theory of computing}, pages 465--474. ACM Press, 2004.

\bibitem{Aldous_min_search}
David Aldous.
\newblock Minimization algorithms and random walk on the {$d$}-cube.
\newblock {\em Ann. Probab.}, 11(2):403--413, 1983.

\bibitem{BBC}
Robert Beals, Harry Buhrman, Richard Cleve, Michele Mosca, and Ronald de~Wolf.
\newblock Quantum lower bounds by polynomials.
\newblock {\em J. ACM}, 48(4):778--797, 2001.

\bibitem{modern_graph_theory}
B{\'e}la Bollob{\'a}s.
\newblock {\em Modern graph theory}, volume 184 of {\em Graduate Texts in
  Mathematics}.
\newblock Springer-Verlag, 1998.

\bibitem{Durr_Hoyer}
Christoph D{\"u}rr and Peter H{\o}yer.
\newblock A quantum algorithm for finding the minimum.
\newblock quant-ph/9607014.

\bibitem{separator_genus}
John~R. Gilbert, Joan~P. Hutchinson, and Robert~Endre Tarjan.
\newblock A separator theorem for graphs of bounded genus.
\newblock {\em J. Algorithms}, 5(3):391--407, 1984.

\bibitem{pls_class}
David~S. Johnson, Christos~H. Papadimitriou, and Mihalis Yannakakis.
\newblock How easy is local search?
\newblock {\em J. Comput. System Sci.}, 37(1):79--100, 1988.

\bibitem{Llewellyn_Tovey_2}
Donna~C. Llewellyn and Craig~A. Tovey.
\newblock Dividing and conquering the square.
\newblock {\em Discrete Appl. Math.}, 43(2):131--153, 1993.

\bibitem{Llewellyn_Tovey_1}
Donna~C. Llewellyn, Craig~A. Tovey, and Michael Trick.
\newblock Local optimization on graphs.
\newblock {\em Discrete Appl. Math.}, 23(2):157--178, 1989.

\bibitem{pap_schaf_yann}
Christos~H. Papadimitriou, Alejandro~A. Scha\"effer, and Mihalis Yannakakis.
\newblock On the complexity of local search.
\newblock In {\em Proceedings of the twenty-second annual ACM symposium on
  Theory of computing}, pages 438--445. ACM Press, 1990.

\bibitem{Miklos_Mario}
Miklos Santha and Mario Szegedy.
\newblock Quantum and classical query complexities of local search are
  polynomially related.
\newblock In {\em Proceedings of the thirty-sixth annual ACM symposium on
  Theory of computing}, pages 494--501. ACM Press, 2004.

\bibitem{zhang}
Shengyu Zhang.
\newblock ({A}lmost) tight bounds for randomized and quantum local search on
  hypercubes and grids.
\newblock quant-ph/0504085.

\end{thebibliography}

\end{multicols}
\end{document}